\newif\ifpreprint

\preprinttrue

\ifpreprint
\documentclass[prl,twocolumn,showpacs,preprintnumbers,amsmath,amssymb]{revtex4}
\else
\documentclass[prl,preprint,showpacs,preprintnumbers,amsmath,amssymb]{revtex4}
\fi

\usepackage{graphicx} 
\usepackage{dcolumn}  
\usepackage{bm}       

\newcommand{\fref}[1]{Figure~\ref{#1}}
\newcommand{\eref}[1]{equation~(\ref{#1})}
\newcommand{\Mainj}{M_{A}^{\rm inj}}
\newcommand{\Mainjs}{M_{A}^{\rm inj*}}
\newcommand{\Maw}{M_{A}^{\rm whistler}}
\newcommand{\Alfven}{Alfv\'{e}n\ }
\newcommand{\tbn}{\theta_{Bn}}
\newcommand{\wps}{\omega_{ps}}
\newcommand{\wcs}{\Omega_{cs}}
\newcommand{\wpe}{\omega_{pe}}
\newcommand{\wce}{\Omega_{ce}}
\newcommand{\wpi}{\omega_{pi}}
\newcommand{\wci}{\Omega_{ci}}
\newcommand{\vs}{v_{s}}

\newcommand{\ve}{v_{e}}
\newcommand{\bs}{\beta_{s}}
\newcommand{\bi}{\beta_{i}}
\newcommand{\be}{\beta_{e}}
\newcommand{\vas}{v_{A,s}}

\newcommand{\Ma}{M_{A}}
\newcommand{\Va}{v_{A}}

\newcommand{\Vs}{V_{s}}

\newcommand{\hyphen}{{\rm -}}

\newcommand{\para}{\parallel}
\newcommand{\ww}{\omega_0}
\newcommand{\kk}{k_0}
\newcommand{\gamgrow}{\gamma_{\rm grow}}
\newcommand{\gamdamp}{\gamma_{\rm damp}}
\newcommand{\gamtot}{\gamma_{\rm tot}}
\newcommand{\phiht}{\tilde{\phi}}
\newcommand{\tlc}{\theta_{lc}}
\newcommand{\Bmax}{B_{\rm max}}

\begin{document}

\title{A Critical Mach Number for Electron Injection in Collisionless Shocks}

\author{Takanobu Amano$^1$}
\email{amanot@stelab.nagoya-u.ac.jp}
\author{Masahiro Hoshino$^2$}
\affiliation{
$^1$Department of Physics, Nagoya University, Nagoya, 464-8602, Japan \\
$^2$Department of Earth and Planetary Science, University of Tokyo,
Tokyo, 113-0033, Japan
}

\date{\today}

\begin{abstract}
Electron acceleration in collisionless shocks with arbitrary magnetic field orientations is discussed. It is shown that the injection of thermal electrons into diffusive shock acceleration process is achieved by an electron beam with a loss-cone in velocity space that is reflected back upstream from the shock through shock drift acceleration mechanism. The electron beam is able to excite whistler waves which can scatter the energetic electrons themselves when the \Alfven Mach number of the shock is sufficiently high. A critical Mach number for the electron injection is obtained as a function of upstream parameters. The application to supernova remnant shocks is discussed.
\end{abstract}

\preprint{}

\maketitle

It is widely believed that a large part of cosmic rays (CRs) in our galaxy are produced by supernova remnant (SNR) shocks. The diffusive shock acceleration (DSA) model, which is based on the idea of particle scattering by plasma waves, has been considered as the most plausible mechanism of CR acceleration at collisionless shocks \cite{1987PhR...154....1B}. Since ions can easily be scattered either by ambient or self-generated magnetohydrodynamic (MHD) waves, the ion acceleration (or injection) is relatively easier to understand \cite{1995A&A...300..605M}. On the other hand, resonant scatterings of thermal electrons are usually less efficient because of their small gyroradii, and thus, mildly relativistic energies are needed for the resonance with MHD turbulence in the typical interstellar medium. This, the so-called injection problem, has been a long-standing unresolved issue in the DSA theory. Indeed, it is known from in situ observations in the heliosphere that the gradual increase of energetic particles toward the shock, as predicted by the DSA theory, is the common feature for ions, while it is not for electrons. The accelerated electrons (typically with energies $\lesssim 10 {\rm keV}$) are usually confined in a much smaller region close to the shock \cite{1989JGR....9410011G,2006GeoRL..3324104O} except for rare occasions \cite{1999Ap&SS.264..481S,2009EP&S...61..603O}. These observations in the heliosphere support the theoretical expectation that the injection of electrons is far less efficient than that of ions. On the other hand, there is strong evidence for the presence of ultra-relativistic electrons in young SNR shocks \cite{1995Natur.378..255K}. The most striking difference between the heliosphere and other astrophysical environments seems to be the difference in \Alfven Mach numbers $\Ma = \Vs / \Va$, where $\Vs$, and $\Va$ are the shock, and the \Alfven velocity, respectively. The solar wind at 1 AU has an average Mach number of $\sim 5\hyphen10$, while it can be much higher at SNR shocks. The purpose of this paper is to show that there exists a critical Mach number above which the electron injection can naturally be explained.

The injection of electrons requires either a mechanism by which thermal electrons are accelerated to mildly relativistic energies, or the generation of high-frequency whistler waves propagating toward the shock which can scatter sub-relativistic electrons. \citet{1992ApJ...401...73L} proposed the injection process through the excitation of high-frequency (oblique) whistler waves by the anisotropy of pre-existing CR electrons. A different approach is to consider the pre-acceleration of thermal electrons to the injection threshold through plasma microinstabilities in the thin shock layer. One possible scenario is to consider lower-hybrid waves excited by the reflected ions in quasi-perpendicular shocks ($\tbn \gtrsim 45$, where the shock angle $\tbn$ is the angle between the shock normal and the upstream magnetic field), while it requires a rather high shock velocity (a few per cent of the speed of light), as well as large magnetic field for the injection \cite{1997MNRAS.291..241M}. Another mechanism working in weakly magnetized plasmas is the shock surfing acceleration (SSA) \cite{2001PhRvL..87y5002M,2002ApJ...572..880H}, which was recently shown to be efficient even in two dimensions \cite{2009ApJ...690..244A,2009PhPl...16j2901A}. \citet{2007ApJ...661..190A} have recently demonstrated using one-dimensional particle-in-cell simulations that suprathermal electrons accelerated by the SSA are subject to further energization by the shock drift acceleration (SDA). The process can actually be a possible solution to the electron injection problem because the accelerated electrons can self-generate \Alfven waves. We begin our discussion by deriving the condition required for the self-generation of \Alfven waves.

\ifpreprint
\begin{figure}[t]
 \includegraphics[scale=1.0]{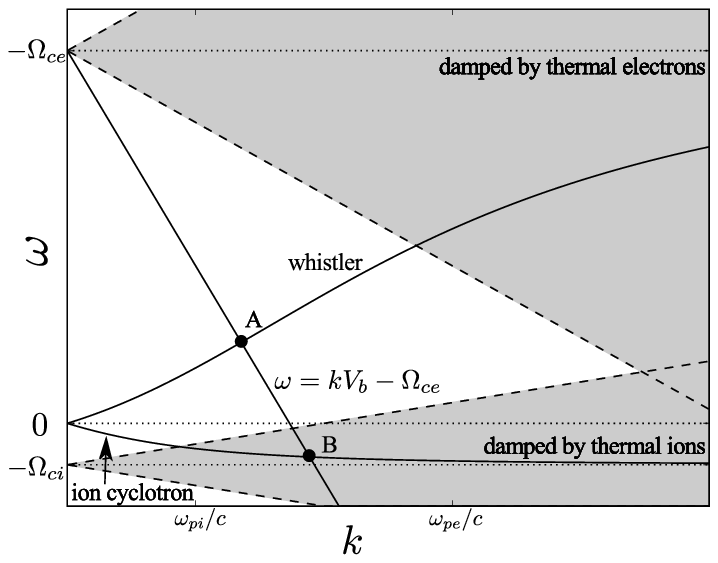}
 \caption{Schematic dispersion diagram for circularly polarized electromagnetic waves in an electron-ion plasma. Positive (negative) frequency corresponds to the right-hand (left-hand) polarization. The cyclotron resonance condition $\omega = k V_b - \wce (V_b < 0)$ is also shown. Waves in the shaded regions are strongly damped by the cyclotron damping of thermal plasma.}
 \label{fig:dispersion}
\end{figure}
\fi

The SDA can be understood as an adiabatic mirror reflection process in the de Hoffman-Teller frame (HTF) where the motional electric field vanishes. The beam velocity of the reflected population measured in the upstream rest frame is given by $V_b \simeq 2 \Vs / \cos \tbn$. We consider electromagnetic instabilities excited by the reflected electron beam. In the following, the mass, charge, density and temperature of particle species $s$ are respectively given by $m_s$, $q_s$, $n_s$, $T_s$. We then define the plasma frequency, gyrofrequency, thermal velocity, \Alfven velocity, and plasma beta as follows: $\wps = \sqrt{4\pi n_s q_s^2/m_s}$, $\wcs = q_s B / m_s c$, $\vs = \sqrt{2 T_s/m_s}$, $\vas = B/\sqrt{4 \pi n_s m_s}$, $\bs = \vs^2/\vas^2$, where $B$ is the ambient magnetic field strength. Note the definition of the gyrofrequency includes the sign of charge (i.e., $\wce < 0$). \fref{fig:dispersion} schematically shows the dispersion relation of circularly polarized electromagnetic waves propagating parallel to the magnetic field in an electron-ion plasma. Note that the positive $k$ direction is pointing toward the shock (thus $V_b < 0$). The beam strongly interacts with the background plasma only at the intersection points between the resonance condition $\omega = k V_b - \wce$ and the normal mode of the plasma (A and B in the figure). If one considers a cold beam, i.e., $\delta$ function in velocity space $f(v_{\para}, v_{\perp}) \propto \delta (v_{\para} - V_b) \delta (v_{\perp})$ ($v_{\para}$, $v_{\perp}$ are the velocity components parallel, and perpendicular to the ambient magnetic field, respectively), the excitation of whistler waves at the point A is prohibited because otherwise the momentum conservation law is violated. On the other hand, the wave growth at the point B should overcome the cyclotron damping of thermal ions, which is strong at short wavelength $k c/\wpi \gtrsim 1$ (where $c$ is the speed of light). Therefore, the excitation of \Alfven waves requires that the reflected electron beam interacts with the normal mode at $k c /\wpi \lesssim 1$. Assuming $|\omega| \ll |\wce|$, we obtain the condition \cite{2007ApJ...661..190A}
\begin{eqnarray}
 \Ma \gtrsim \frac{\cos \tbn}{2} \frac{m_i}{m_e}.
  \label{eq:critical-alfven}
\end{eqnarray}
The required Mach number is fairly high especially at quasi-parallel shocks. Although some young SNRs may satisfy the condition, it seems to be too stringent to explain available astrophysical observations.

We now want to relax the requirement by exploring the possibility of the whistler wave excitation at the point A. It is actually possible by considering an electron beam with a loss cone in velocity space. The loss-cone distribution is a natural consequence of the mirror reflection because particles having large pitch angles are preferentially reflected by the magnetic mirror. It can be unstable against the excitation of whistler waves, however, the instability is tend to be suppressed by the cyclotron damping of thermal electrons. The damping of whistler waves is significant for waves satisfying $-k \ve - \wce \lesssim \omega \lesssim + k \ve - \wce$ (the upper shaded region in \fref{fig:dispersion}). Therefore, the wave generation is expected only when the beam velocity is faster than the electron thermal velocity $|V_b| \gtrsim \ve$. The condition can be rewritten as
\begin{eqnarray}
 \Ma \gtrsim \frac{\cos \tbn}{2} \sqrt{\frac{m_i}{m_e} \be} \equiv \Mainj.
  \label{eq:critical-whistler}
\end{eqnarray}
Since the required Mach number is now proportional to $\sqrt{m_i/m_e}$, it is smaller than \eref{eq:critical-alfven} by a factor of $\sim 43/\sqrt{\be}$, thereby greatly relaxes the requirement for plasmas with $\be \sim 1$.

The above analytical expression gives only a rough estimate. While it is useful, we want to determine the condition more precisely. To do so, we need to model a velocity distribution function of the reflected electron beam. We here adopt a cold ring-beam distribution of the form $f(v_{\para}, v_{\perp}) = n_b / (2 \pi V_r) \delta (v_{\para} - V_b) \delta (v_{\perp} - V_r)$ where $n_b$, $V_r$ are the beam density, and ``ring velocity'', respectively. Although it may not necessarily be a realistic distribution, the essential physics is retained for our purpose. The cold plasma dispersion relation for right-hand circularly polarized electromagnetic waves in the presence of the cold ring-beam electrons can be written as
\begin{eqnarray}
 D &=& 1 - \frac{k^2 c^2}{\omega^2}
  - \frac{\wpi^2}{\omega(\omega + \wci)}
  - (1 - \eta) \frac{\wpe^2}{\omega(\omega + \wce)} \nonumber \\
 &-& \eta \frac{\wpe^2}{\omega^2}
  \left[
   \frac{1}{2} \frac{k^2 V_r^2}{(\omega - k V_b + \wce)^2} +
   \frac{\omega - k V_b}{\omega - k V_b + \wce}
  \right] = 0,
  \nonumber
\end{eqnarray}
where $\eta = n_b/n_i  \ll 1$ is the density of the beam normalized to the total density. We seek an approximated unstable solution of the dispersion relation in the low frequency regime ($k c / \omega \gg 1$). It is easy to understand that the contribution from the beam becomes significant when $\omega \simeq \ww \equiv k V_b - \wce$. Thus, we retain only the term proportional to $1/(\omega - \ww)^2$ for the beam. We define $\kk$ as the wavenumber at $\omega = \ww$ on the whistler mode branch in the absence of the beam. We then write $\omega = \ww + \delta \omega$ and expand the dispersion relation in powers of $\delta \omega$ under the assumption $|\delta \omega| \ll |\ww|$. The growth rate $\gamgrow = {\rm Im} (\delta \omega)$ is then obtained as
\begin{eqnarray}
 \frac{\gamgrow}{\wci} \simeq \frac{\sqrt{3}}{2}
  \left(
   \chi \eta \frac{m_i}{m_e} \frac{\kk^2 V_r^2}{2 \wci^2}
  \right)^{1/3},
  \nonumber
\end{eqnarray}
where $\chi = - (\ww+\wci)^2 (\ww+\wce)^2 / (\wci (\wci(\ww+\wce)^2 + \wce(\ww+\wci)^2) m_i/m_e)$ is a positive numerical factor of order unity \citep[see, e.g.,][]{1986JGR....91.1529S}. It is important to note that the growth rate is proportional to $V_r^{2/3}$, in addition to the well-known dependence for beam-plasma instabilities $\eta^{1/3}$. The wave growth, in reality, competes with the the cyclotron damping by a thermal plasma. The cyclotron damping rate $\gamdamp$ for a wave $(\ww, \kk)$ may be written as
\begin{eqnarray}
 \frac{\gamdamp}{\wci} \simeq
  - \frac{\sqrt{\pi}}{\ww \partial D_r / \partial \omega|_{\omega=\ww}}
  \frac{c^2}{\Va^2}
  \sum_{s=i,e} \frac{n_s}{n_i} \frac{m_i}{m_s} \frac{\wci}{\kk \vs}
  e^{- \left(\frac{\ww + \wcs}{\kk \vs} \right)^2}
 \nonumber
\end{eqnarray}
where $D_r$ is the real part of $D$. For calculation of the damping rate, the contribution from the beam component is ignored. We assume that the ion temperature is the same as the electrons ($\bi= \be$), and consider the damping rate as a function of $\be$ for a given $(\ww, \kk)$. Since $(\ww, \kk)$ depends only on $V_b$, the total growth rate $\gamtot = \gamgrow + \gamdamp$ is solely determined by $V_b$, $V_r$, $\be$, and $\eta$. In the present paper, the beam density $\eta$ is considered as a free parameter, on which the dependence of the result should be investigated.

For our purpose, it is sufficient to write the ring velocity $V_r$ by using upstream quantities. It may be estimated as the minimum perpendicular velocity required for the mirror reflection
\begin{eqnarray}
 V_r^2 \simeq \left( V_b^2 + \frac{m_i}{m_e} \phiht \Vs^2 \right) \tan^2 \tlc,
  \nonumber
\end{eqnarray}
where $\phiht = 2 e \phi^{\rm HTF} / m_i \Vs^2$, and $\tlc$ are the normalized cross-shock electrostatic potential measured in the HTF, and the loss-cone angle, respectively. Since the potential is estimated to be $\phiht \simeq 0.1\hyphen0.2$ from in situ observations \cite{1988JGR....9312923S}, a fixed value of $\phiht = 0.2$ is used. The loss-cone angle $\tlc$ is defined as $\tan^2 \tlc = B_0 / (\Bmax - B_0)$ using the magnetic field strength in the upstream $B_0$ and at the maximum $\Bmax$. It is worth noting that the compressed transverse magnetic field reaches its maximum at the so-called overshoot region, where the compression ratio $r$ is greater than expected from Rankine-Hugoniot relations. Since the foreshock region of quasi-parallel shocks are always accompanied by large amplitude MHD turbulence, it is natural to assume the presence of fluctuating transverse magnetic fields $\delta B \sim B_0$ ahead of the shock. Hence, we may estimate the maximum transverse magnetic field as $\sim \delta B + r \sin \tbn B_0$. By assuming $\delta B = B_0$, we obtain the maximum magnetic field strength $(\Bmax/B_0)^2 =  \cos^2 \tbn + (r \sin \tbn + 1)^2$. Thus, we obtain a finite compression ratio even at a purely parallel shock ($\tbn = 0$). We assume $r = 6$ in the following discussion.

Since both $V_b$ and $V_r$ are written in terms of $\Ma$ and $\tbn$, the total growth rate $\gamtot$ is now a function of $\Ma$, $\tbn$, and $\be$ for a given $\eta$. We define a critical Mach number $\Mainjs$ for the electron injection as a numerical solution of $\gamtot = 0$. \fref{fig:critical} shows the solutions as functions of $\tbn$ for $\eta = 10^{-3}$, $10^{-4}$, $10^{-5}$ , where a fixed $\be = 1$ is used. For comparison, a solution with a constant $\tan \tlc$ assuming $\Bmax = r B_0$ is also shown by a dashed line for the case of $\eta = 10^{-4}$. These two curves are almost identical in the quasi-perpendicular regime, while the deviation becomes larger at quasi-parallel shocks. We find that the result only weakly depends on the choice of $\phiht$, $\tlc$ at $\tbn \gtrsim 45$ for the parameter range of our interest. On the other hand, the result may depend on the assumption at a quasi-parallel shock due to its intrinsic complexity. For instance, our model expects large ratios of $V_r/V_b$ for quasi-parallel shocks (e.g., $V_r/V_b \simeq 4.7$ at $\tbn = 30$) compared to quasi-perpendicular shocks ($V_r/V_b \lesssim 1$), which makes it easier to excite whistler waves. Although this trend itself is naturally expected, specific quantities will certainly depend on the shock structure. Therefore, we think that our estimate in the quasi-parallel regime includes an uncertainty factor of order unity, which can be studied only by using self-consistent numerical simulations.

\ifpreprint
\begin{figure}[t]
 \includegraphics[scale=1.0]{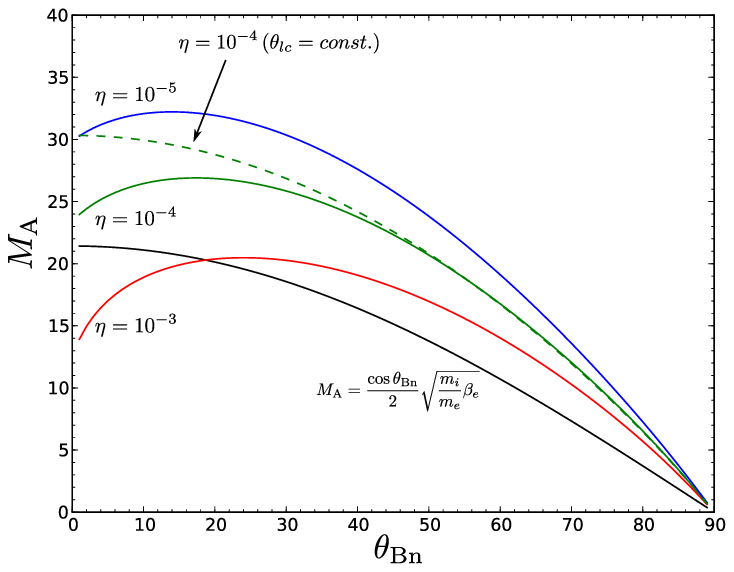}
 \caption{Critical Mach number for the electron injection as a function $\tbn$ for the case $\be = 1$. $\eta = n_b/n_i$ is the normalized beam density. The dashed line corresponds to the case with a constant $\tlc$. The black solid line shows the analytical expression given in \eref{eq:critical-whistler} for reference.}
 \label{fig:critical}
\end{figure}
\fi

The above result well explains a recent statistical analysis of the Earth's bow shock crossings observed by the Geotail satellite \cite{2006GeoRL..3324104O}. \citet{2006GeoRL..3324104O} claimed that the power-law index of accelerated electrons $\Gamma$ clearly depends on the so-called whistler critical Mach number $\Maw = (\alpha/2) \cos \tbn \sqrt{m_i/m_e}$, which is defined as the critical point above which whistler waves cannot propagate upstream. Note that $\alpha$ is a numerical factor of order unity, and they used $\alpha = \sqrt{27/16} \simeq 1.3$. It is easy to recognize that $\Maw$ is close to $\Mainj$ in plasmas with $\be \sim 1$, which is statistically valid in the averaged solar wind. We see from their Figure 4(b) that $\Gamma$ flattens when Mach numbers go slightly above $\Maw$. In a parameter regime relevant to their observations (i.e., $10^{-5} \lesssim \eta \lesssim 10^{-3}$, $\be \sim 1$, and $\tbn \gtrsim 60$), we find roughly $\Mainjs/\Mainj \sim 2$ or slightly below. The result is consistent with the observation in that the critical Mach number for the efficient electron acceleration is larger than $\Maw$ by a factor of $\sim 2$. Note that our extensive parameter survey shows that roughly $\Mainjs/\Mainj \sim 1 {\rm -} 3$ for a more wide range of parameters.

Since the growth rate of whistler waves well above the threshold is much larger than the ion gyrofrequency (e.g., $\gamtot/\wci \simeq 21$ for $\Ma = 10$, $\be = 1$, $\tbn = 80$, and $\eta = 10^{-4}$), the spatial scale of the wave growth is much smaller than the convective gyroradius of ions $\Vs / \gamtot \ll \Vs/\wci$, which represents the typical width of the quasi-perpendicular shock transition region. Therefore, we expect that the electron acceleration occurs within the thin shock transition region, which again well explains the observed characteristics of energetic electrons. We should note that loss-cone type distributions are frequently observed in the foreshock region and thought to be a result of the mirror reflection process  \cite{1990JGR....95.4155F}. Therefore, the assumption of loss-cone distributions seems to be reasonable at first glance. On the other hand, the presence of a loss cone in the upstream region means the absence of whistler-wave generation and resulting pitch-angle scattering, which is most likely due to lower Mach numbers. Careful reanalysis of the electron measurements as a function of Mach numbers within or immediate upstream of the shock is needed to further ensure the applicability of the present theory. It is important to mention that \citet{1989JGR....9410011G} reported that suprathermal electrons are observed primarily around quasi-perpendicular shocks, with an exception of an unusually high Mach number solar wind $\Ma \sim 41$. According to the authors, the shock ($\tbn \sim 45$) was quasi-parallel in character but has an enhanced suprathermal electron flux, which is also consistent with our prediction.

So far, we have confirmed that the present theory is strongly supported by in situ measurements of the bow shock. It is interesting that the critical parameter $\Ma/\sqrt{\be}$ is independent of both the ambient density and the magnetic field strength. Our result indicates that the required condition for the electron injection may be written by using typical SNR shock parameters:
\begin{eqnarray}
 \frac{\Ma}{\sqrt{\be}} \simeq 68 \left(\frac{\Vs}{3000\,{\rm km/s}}\right)
  \left(\frac{T_e}{10\,{\rm eV}}\right)^{-1/2} \gtrsim 30 \label{eq:threshold}
\end{eqnarray}
regardless of the upstream magnetic field directions ($\tbn$).
Therefore, we conclude that the injection and subsequent acceleration of electrons through DSA process will operate in young SNR shocks. This gives, for the first time, the most natural explanation for the difference in the electron acceleration efficiencies observed in different environments.

In the present study, we have performed only the linear instability analysis. However, it is known that whistler waves undergo both forward and inverse cascade processes \cite{1972PhRvL..29..249F}, which may enhance the efficiency of injection and further acceleration to relativistic energies. We should note that a quantitative estimate of the injection rate $\eta \sim 10^{-4}$ is given by \cite{2007ApJ...661..190A} with some simplifying assumptions. Although it is possible to consider nonlinear processes affect the estimate quantitatively, the injection process itself will not be modified as it has only a weak dependence on the injection rate. Therefore, we think that the present mechanism for the electron injection is robust, and there is no reason why the process should not work in astrophysical shocks.

One of the most important remaining problems, when it applies to SNRs, is probably the effects of CR back reaction. It is believed that CRs amplify the upstream magnetic field up to $\sim 0.1{\rm -}1 \, {\rm mG}$ as suggested by X-ray observations of some young SNRs. While the condition given by \eref{eq:threshold} is independent of the magnetic field strength, other effects such as pre-shock electron heating and deceleration of the upstream plasma, if strong, might become important. Although we think the injection mechanism itself will remain unchanged because of the locality of the process, it is interesting to investigate the consequence of such nonlinear shock behaviors to the electron injection efficiency.

\begin{acknowledgments}
This work was supported by the Global COE Program of Nagoya University ``Quest for Fundamental Principles in the Universe (QFPU)'' from JSPS and MEXT of Japan. Part of this work was done at the KITP in Santa Barbara.
\end{acknowledgments}



\ifpreprint
\else
\begin{figure}[h]
 \includegraphics[scale=1.0]{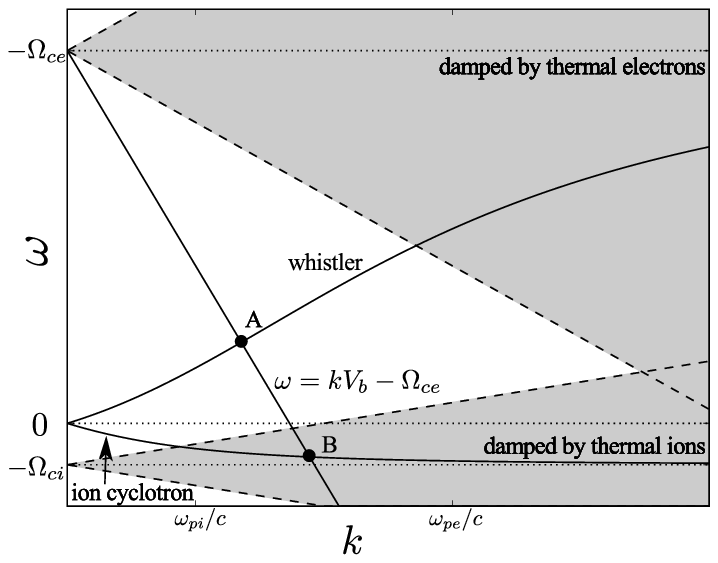}
 \caption{Schematic dispersion diagram for circularly polarized electromagnetic waves in an electron-ion plasma. Positive (negative) frequency corresponds to the right-hand (left-hand) polarization. The cyclotron resonance condition $\omega = k V_b - \wce (V_b < 0)$ is also shown. Waves in the shaded regions are strongly damped by the cyclotron damping of thermal plasma.}
 \label{fig:dispersion}
\end{figure}

\begin{figure}[h]
 \includegraphics[scale=1.0]{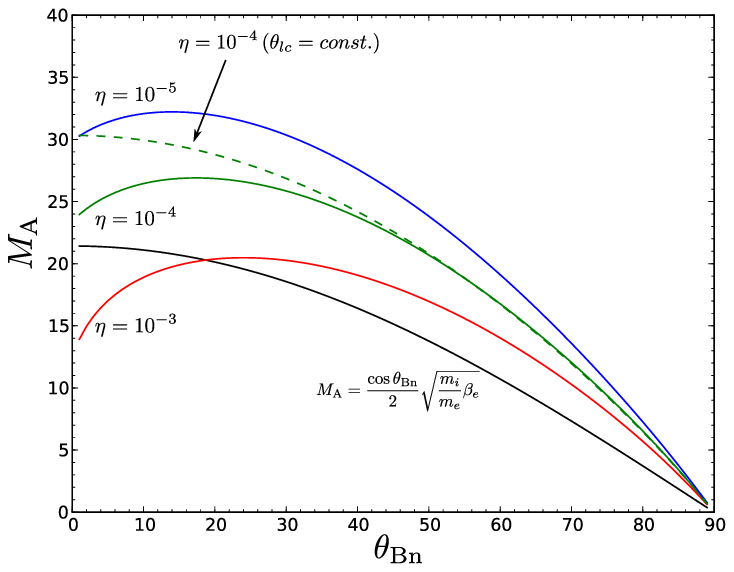}
 \caption{Critical Mach number for the electron injection as a function $\tbn$ for the case $\be = 1$. $\eta = n_b/n_i$ is the normalized beam density. The dashed line corresponds to the case with a constant $\tlc$. The black solid line shows the analytical expression given in \eref{eq:critical-whistler} for reference.}
 \label{fig:critical}
\end{figure}
\fi

\end{document}